\documentclass{vgtc}                          

\ifpdf
  \pdfoutput=1\relax                   
  \usepackage{graphicx}                
  \DeclareGraphicsExtensions{.pdf,.png,.jpg,.jpeg} 
\else
  \usepackage{graphicx}                
  \DeclareGraphicsExtensions{.eps}     
\fi%

\graphicspath{{figures/}{pictures/}{images/}{./}} 

\usepackage{microtype}                 
\PassOptionsToPackage{warn}{textcomp}  
\usepackage{textcomp}                  
\usepackage{mathptmx}                  
\usepackage{times}                     
\usepackage{cite}                      
\usepackage{tabu}                      
\usepackage{booktabs}                  
\usepackage{multirow}
\usepackage{subfigure}

\onlineid{0}

\vgtccategory{Research}

\vgtcinsertpkg

\title{Auditory Feedback for Standing Balance Improvement in  Virtual Reality}

\author{M. Rasel Mahmud\thanks{e-mail: m.raselmahmud1@gmail.com}\\ %
     \parbox{1.4in}{\scriptsize \centering Computer Science \\ The University of Texas at San Antonio} %
\and Michael Stewart\thanks{e-mail: michael.stewart@utsa.edu }\\ %
     \parbox{1.4in}{\scriptsize \centering Kinesiology \\ The University of Texas at San Antonio} %
\and Alberto Cordova\thanks{e-mail: Alberto.Cordova@utsa.edu}\\ %
     \parbox{1.4in}{\scriptsize \centering Kinesiology \\ The University of Texas at San Antonio} %
\and John Quarles\thanks{e-mail: John.Quarles@utsa.edu}\\ %
     \parbox{1.4in}{\scriptsize \centering Computer Science \\ The University of Texas at San Antonio}}

\abstract{
Virtual Reality (VR) users often experience postural instability, i.e., balance problems, which could be a major barrier to universal usability and accessibility for all, especially for persons with balance impairments. Prior research has confirmed the imbalance effect, but minimal research has been conducted to reduce this effect. We recruited 42 participants (with balance impairments: 21, without balance impairments: 21) to investigate the impact of several auditory techniques on balance in VR, specifically spatial audio, static rest frame audio, rhythmic audio, and audio mapped to the center of pressure (CoP). Participants performed two types of tasks - standing visual exploration and standing reach and grasp. Within-subject results showed that each auditory technique improved balance in VR for both persons with and without balance impairments. Spatial and CoP audio improved balance significantly more than other auditory conditions. The techniques presented in this research could be used in future virtual environments to improve standing balance and help push VR closer to universal usability.
} 

\keywords{Virtual Reality, balance, postural stability, auditory feedback, VR accessibility, VR usability, Head-Mounted Display}

\CCScatlist{
  \CCScatTwelve{Human-centered computing}{Visu\-al\-iza\-tion}{Visu\-al\-iza\-tion techniques}{Virtual Reality};
  \CCScatTwelve{Human Computer
Interaction (HCI)}{Visualization design and evaluation methods}{User studies}{}
}

\begin{document}


\firstsection{Introduction}

\maketitle

Over one billion people (15\%) around the world have disabilities \cite{WinNT}. Virtual reality (VR) is not accessible to many persons with disabilities \cite{agrawal2009disorders,ferdous2016visual,ferdous2018investigating,guo2013effects,samaraweera2013latency}. Unfortunately, these populations are rarely consulted during VR research and development, which can result in noninclusive and inaccessible experiences. For example, persons with balance impairments (BI) may not be able to stand safely in VR experiences. This limitation may prevent users from engaging in parts of the VR experience. In addition to affecting persons with BI, VR Head Mounted Displays (HMDs) also significantly disrupt the balance of users without BI \cite{epure2014effect,lott2003effect,robert2016effect}. However, minimal research has been conducted to reduce this effect. If these imbalance issues could be mitigated, both persons with and without BI could more readily benefit from consumer VR applications (e.g., education, physical fitness, and entertainment). 

In assistive technology research, there have been numerous feedback techniques of various modalities \cite{franco2012ibalance,sienko2017role} created to help improve balance in reality. For example, researchers have used vibrotactile feedback to improve the balance of persons with low vision \cite{velazquez2010wearable}. Visual feedback also has been used to improve balance \cite{thikey2011need,vcakrt2010exercise,sutbeyaz2007mirror}
. However, there has been minimal research into how auditory feedback can affect balance in VR. 

In this research, we conducted empirical studies in which participants (i.e., participants with and without BI) attempted to maintain balance while standing in VEs with various approaches to auditory feedback (i.e., spatial, static rest frame, rhythmic, and audio based upon Center of Pressure (CoP)). The purpose of this work is to make immersive VR more accessible for all persons. The results are intended to give future VR developers an understanding of how auditory feedback can be applied to improve the accessibility of VR.

\section{Background Study}
\subsection{Imbalance in VR}
Previous literature has reported that there is an imbalance effect caused by virtual environments (VEs). In the early 2000s, a finding reported that postural control mechanism was affected in VEs, which could cause imbalance and motion sickness \cite{takahashi2001change}. Participants had less control on balance in the virtual environment than the real environment \cite{kelly2008visual}. HMDs also provided less stabilization of balance in the virtual environment than the real environment \cite{kelly2019visual}. As HMDs block out visual feedback from the real world, participants wearing HMDs may lose balance because of end-to-end latency and illusory sensation of body movement caused by VEs \cite{soltani2020influence, martinez2018analysing}. Longer immersion in VR environments also caused postural instability \cite{murata2004effects}. 
Other works have also investigated the imbalance effects on gait (walking patterns) \cite{sondell2005altered}. Thus, balance issues are a known problem for users in HMD-based VR. However, there have been few previous efforts to mitigate this problem. This inspired us to investigate solutions to improve balance in VR.

Although most previous research was focused on users without disabilities \cite{lott2003effect,epure2014effect,robert2016effect,horlings2009influence}, research from Ferdous et al. \cite{ferdous2018investigating} was one of the few that has investigated balance in VR for persons with multiple sclerosis (MS). They studied how different visual components (frame rate, field of view, display resolution) affected balance in VR for both persons with and without BI. They reported that postural instability increases significantly with the decrease of frame rate and field of view for the participants with BI, but no effect of display resolution on balance was found. On the other hand, they did not find any effect of any visual components on balance for the participants without BI. We have replicated their visual exploration task in our current study.

VR has a long history of enabling effective rehabilitation approaches for improving balance and gait \cite{de2016effect,bisson2007functional,meldrum2012effectiveness,park2015effects,cho2016treadmill,rendon2012effect,duque2013effects, bergeron2015use}. However, most of this work has either not used HMDs or was constrained to seated tasks. Thus, the effects of immersive VR with HMDs on standing balance have minimally been studied, which motivated us to investigate the standing balance in VR environments with HMDs for persons with disabilities.

\subsection{Assistive Technology: Auditory Feedback for Balance Improvement in the Real World}
 Although they have received less attention than visual feedback approaches \cite{gandemer2016sound}, previous studies conducted in non-VR settings reported that auditory cues could significantly improve postural control in the real world. For example, auditory feedback based upon the lean of the user helped to correct posture for participants without BI \cite{chiari2005audio}. \textit{Spatial audio} - audio that the user can localize in 3D - was found to be effective in maintaining balance for participants with BI \cite{stevens2016auditory}. Ross et al. reported auditory white noise - an \textit{auditory static rest frame} - decreased postural sway for participants who were over the age of 65 \cite{ross2016auditory}. Hasegawa et al. \cite{hasegawa2017learning} reported auditory feedback based on \textit{center of pressure} (CoP) - the center pressure point of the supporting surface, typically measured with multiple pressure sensors on a balance board or force plate. The authors increased the pitch of the sound when the CoP displacement was in the backward direction and decreased the pitch of the sound when the CoP displacement was in the forward direction, which improved balance for participants without BI. Also, \textit{rhythmic audio} - hearing a steady beat -  improved gait in individuals with neurological problems (e.g., multiple sclerosis, Parkinson’s) and elderly persons \cite{ghai2018effect}.  However, among the different auditory techniques, spatial audio was more heavily used as it is intended to be more realistic and natural\cite{chong2020audio, pinkl2020spatialized}. These previous studies motivated us to investigate whether spatial, CoP, rhythmic, static rest frame audio could improve balance significantly for both participants with and without BI.

Very few prior studies investigated auditory techniques in a VR environment for improving balance. For example, Gandemer et al. investigated the postural sway of blindfolded people and reported that spatial audio in an immersive virtual environment improved postural stability \cite{gandemer2017spatial}. Spatial audio was generally preferred to be used in VR in most of the cases as it provided greater immersion \cite{wenzel2017perception,naef2002spatialized}. However, auditory feedback-based balance improvement in an immersive VR environment has been minimally explored, which motivated us to investigate the effect of different auditory feedback (spatial, static rest frame, rhythmic, and CoP) on balance in the VR environment.

%

\section{Methods}
\subsection{Hypotheses}
Our study investigated the effect of different auditory conditions (spatial, static rest frame, rhythmic, and CoP) on balance in VR environments. Based on the knowledge from the prior assistive technology and VR balance literature, we investigated the following hypotheses:

H1: Each VR-based auditory condition (spatial, static rest frame, rhythmic, and CoP) will improve balance significantly more than the no audio in VR condition.

H2: Spatial auditory feedback may facilitate better balance than other auditory feedback techniques.

H3: Postural stability will decrease in no audio in VR condition compared to the baseline (non-VR) condition.

\subsection{Participants, Selection Criteria, and Screening Process}
We recruited 42 participants (Male:11, Female:31) from the San Antonio area to investigate the effect of auditory feedback on balance in VR. Twenty-one participants (Male:7, Female:14) aged 18-75 had BI due to multiple sclerosis (MS) (18 participants), age (1 participant), arthritis (1 participant), and vestibular dysfunction (1 participant). The remaining twenty-one participants (Male:4, Female:17) were without BI and had no MS, arthritis, vestibular dysfunction, or any other physical issues, but they were of similar age, weight, and height to that of the participants with BI. There were 52.4\%  White, 28.6\% Hispanic, 23.8\% African American in participants with BI. For persons without BI, there were 19\% White, 52.4\% Hispanic, 28.6\% African American, 9.5\% American Indian, 9.5\% Asian. The mean (SD) age, height, weight, and gender details for both groups (with and without BI) have been shown in Table 1. We had more female participants than males in our study. This is because we recruited from the population with MS, which is statistically more common in females \cite{WinNT}. All participants were able to walk without assistance.  Participants were recruited from local MS support groups, local rehabilitation hospitals, and church communities in the local area. Verbal recruitment from the authors, email lists, websites, and flyers were the primary means for recruiting.

\paragraph{Screening Process:} First, we interviewed every potential participant by phone to verify if they were qualified for this study. For example, we asked them some simple questions first, including the year and date (to loosely assess mental faculties) and demographic information. We did not select any person who could not understand the questions or did not have English language fluency. Then we asked them about the causes of their balance problems if known. We also ensured that participants were demographically similar (i.e., age, height, weight) across populations. We excluded the participants from the study who were on medication to improve their balance or could not stand without assistance.

\begin{table}[]
\caption{Descriptive statistics for participants}
    \label{tab:my_label}
\setlength{\tabcolsep}{1pt}
\begin{tabular}{|c|cc|cc|cc|cc|}
\hline
\multirow{0}{*}{}{\textbf{\begin{tabular}[c]{@{}c@{}}Participant \\ Group\end{tabular}}} & \multicolumn{2}{c|}{\textbf{Participants}} & \multicolumn{2}{c|}{\textbf{Age (years)}} & \multicolumn{2}{c|}{\textbf{Height (cm)}} & \multicolumn{2}{c|}{\textbf{Weight (kg)}} \\ \cline{2-9} 
                                                                                       & \multicolumn{1}{c|}{Male}     & Female     & \multicolumn{1}{c|}{Mean}      & SD       & \multicolumn{1}{c|}{Mean}      & SD       & \multicolumn{1}{c|}{Mean}      & SD       \\ \hline
\textbf{BI}                                                                            & \multicolumn{1}{c|}{7}        & 14         & \multicolumn{1}{c|}{46.5}      & 13.0     & \multicolumn{1}{c|}{164.84}    & 12.62    & \multicolumn{1}{c|}{82.79}     & 22.18    \\ \hline
\textbf{Without BI}                                                                    & \multicolumn{1}{c|}{4}        & 17         & \multicolumn{1}{c|}{43.2}      & 12.6     & \multicolumn{1}{c|}{164.33}    & 12.7     & \multicolumn{1}{c|}{85.25}     & 17.96    \\ \hline
\end{tabular}
\end{table}

\subsection{System Description}
We used the following equipment for our study.
\paragraph{Balance Measurement:} BTrackS Balance Plate was used to measure participants’ balance in each condition. The sampling frequency of the balance plate was 25 HZ, which would yield a total of 500 data points in a 20-second trial, for example.

\paragraph{Safety Equipment:} A harness supported all participants to prevent them from sudden falls. The harness was attached to a partial weight-bearing suspension system. Both the harness and the suspension system were from Kaye Products Inc.

\paragraph{Computers, VR Equipment, and Software:} We designed the VEs in Unity3D. The HTC Vive had a 2160 x 1200 pixel resolution with a refresh rate of 90 Hz and a 110-degree field of view. We used integrated HMD headphones for hearing different audios for our study. Vive controllers were used for the standing reach and grasp task. We used a computer in this study to render the VE and record the data. The system was equipped with Intel Core i7 Processor (4.20 GHz), 32 GB DDR3 RAM, NVIDIA GeForce RTX 2080 graphics card, and a Windows 10 operating system. We used the  NI LabView software (v. 2020) to gather data from the BTrackS Balance Plate and streamed the data to Unity3D through sockets. 

\paragraph{Environment:} Our lab had sufficient open space ($>$ 600sq ft.) in a temperature-controlled environment. Only the participant and experimenters were allowed in the lab for the duration of the study. 

\begin{figure}
\subfigure{\includegraphics[width=3cm,height=7.4cm]{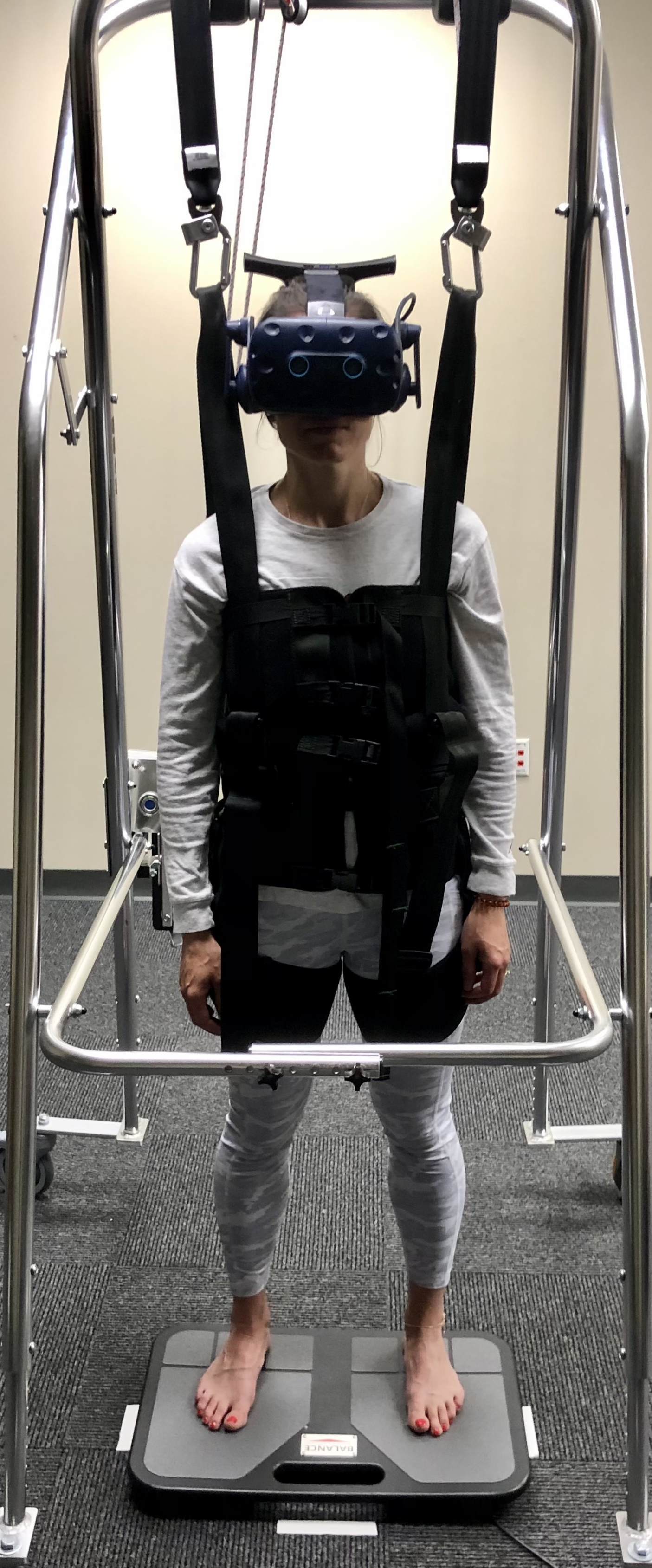}}
\subfigure{\includegraphics[width=5.4cm,height=7.4cm]{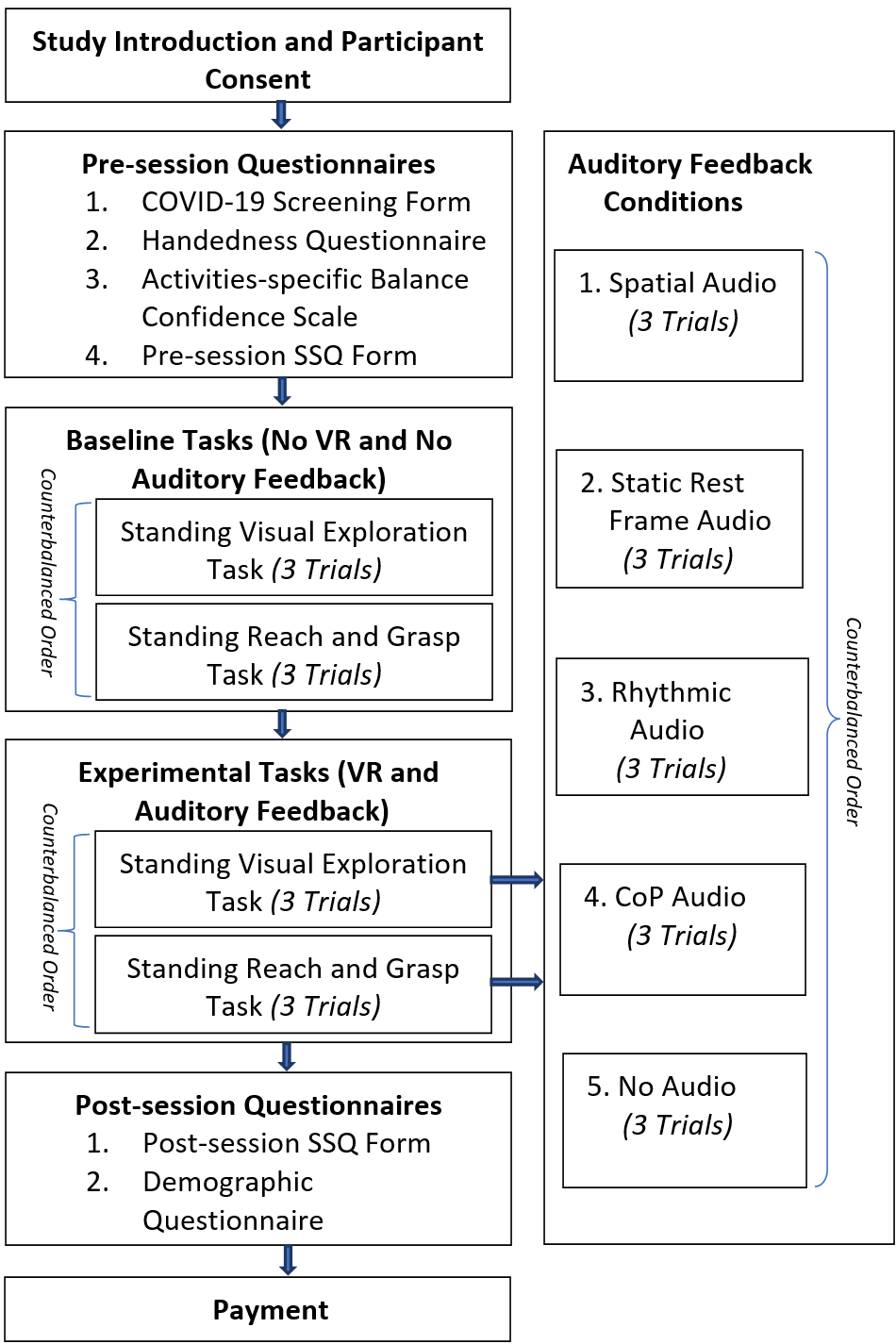}}
\caption{The participants were supported by a harness while they performed the standing visual exploration task (VR environment), standing on balance board and wearing the HMD (left) and study procedure (right).}
\end{figure}


\subsection{Study Conditions}
We investigated four types of VR-based auditory feedback techniques and a condition with no audio to examine how auditory feedback can affect balance in VR. We used white noise for auditory feedback instead of music or user-selected audio tones because white noise has been shown to change the signal-to-noise ratio and improve performance due to the stochastic resonance phenomena \cite{helps2014different}. Auditory white noise was also reported to be effective in reducing postural sway in many previous non-VR studies \cite{sacco2018effects,zhou2021effects,ross2015auditory,harry2005balancing}.

\subsubsection{Spatial Audio} This was 3D audio played in headphones. Specifically, we played spatialized white noise from Unity3D so that as the user turned their head, the noise played at different volumes in each ear to simulate the real stationary audio source. We used Google resonance audio SDK in Unity for audio spatialization as the plugin uses head-related transfer functions (HRTFs) and thus more realistically models 3D sound than the unity default \cite{chong2020audio, pinkl2020spatialized}. The X, Y, and Z coordinates of the 3D audio source in the VE relative to the participant's head were -2.45, -1.43, -1.76. We based this on the known facts that auditory feedback plays a role in balance in the real world \cite{stevens2016auditory,gandemer2017spatial}.    

\subsubsection{Static Rest Frame Audio} This is white noise played in headphones. It had no relation to the listener’s position. This technique was also reported in prior non-VR studies to improve balance in elderly individuals \cite{ross2016auditory}. 

\subsubsection{Pitch and stereo pan feedback on Center of Pressure (CoP)} We played white noise in headphones, similar to static rest frame audio, but the pitch and stereo pan changed based on the center of pressure path obtained from the balance board. In Unity3D, we mapped the pitch to the center of pressure from x coordinate of the balance plate and stereo pan to the center of pressure from y coordinate of the balance plate \cite{hasegawa2017learning}.

\subsubsection{Rhythmic Audio} We played a white noise beat at every 1-second interval. Previous literature suggested that hearing a steady beat can improve balance and gait in both individuals with neurological problems and in the elderly in non-VR environments\cite{ghai2018effect}.

\subsubsection{No Audio} We used this to measure participant's balance in VR with no auditory feedback. Participants still wore the headphones to make it consistent with other conditions, but no audio was played.

\subsection{Auditory Feedback Setup}
We attached the audio to sound sources in Unity3D and modified them based on our study conditions. For each auditory feedback condition, we had a different Unity scene. When the user was ready, we ran the corresponding scene at the beginning of each condition to activate the auditory feedback. We played the scenes in counterbalanced order for all participants to reduce the learning effect. The audio was played through the integrated headphones in the wireless HMD at the start of both tasks. Audio volume was adjusted to the level that the participant felt  was comfortable. 

\subsection{Study Procedure}
The study was approved by Institutional Review Board (IRB). We sterilized all pieces of equipment (e.g., HMD, controllers, balance board, objects, harness, and suspension system) before each user study. At the beginning of the study, the participants filled out a COVID-19 screening questionnaire form, and their body temperature was measured. Then the participant read and signed a consent form. The participants answered handedness questions, which we used to determine their dominant and non-dominant hands \cite{coren1993lateral}. Then we described the whole study procedure. Next, the participants were attached to the harness and the suspension system. The participants were supported by the harness, stood on balance board, and were barefoot for the entire study session. Fig. 1 shows the participant completing the standing visual exploration task (left) and study procedure (right).


\subsubsection{Pre-Session Questionnaires} At the beginning of the study, the participants filled out an Activities-specific Balance Confidence (ABC)\cite{schepens2010short}, and a Simulator Sickness Questionnaire (SSQ)\cite{kennedy1993simulator}.

\subsubsection{Tasks} 
Participants performed a standing visual exploration task and a standing reach and grasp task. Tasks were conducted in both a VR environment and a non-VR environment. The VEs were the replications of the real environments. We ensured that both tasks were performed in a counterbalanced order to mitigate the possibility of confounding variables such as the learning effect.

\begin{figure}
\subfigure{\includegraphics[width=4.25cm,height=4.75cm]{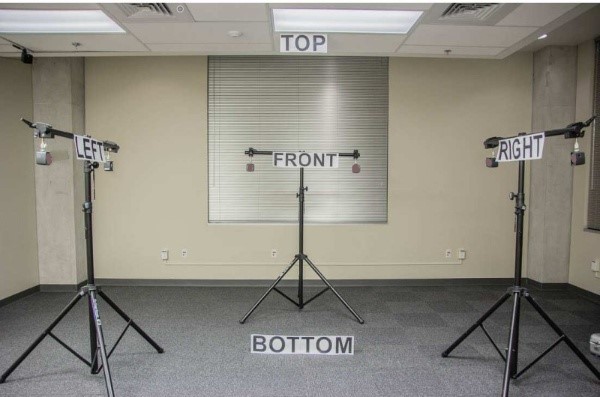}}
\subfigure{\includegraphics[width=4.25cm,height=4.75cm]{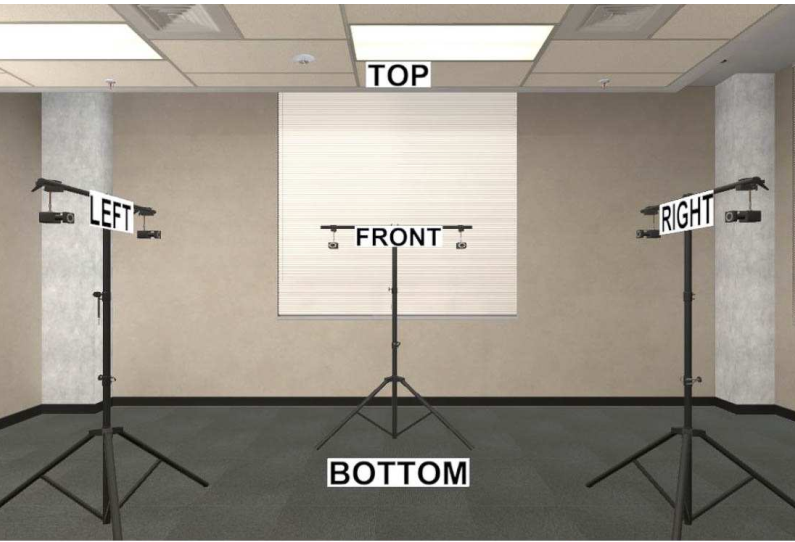}}
\caption{Comparison between real environment (left) and virtual environment (right) for standing visual exploration task.}
\end{figure}


\paragraph{Standing Visual Exploration Task:} A pre-recorded instruction was played at the beginning of each condition and trial which directed participants to look at markers placed at different locations throughout the room. The markers had directional labels, and the instruction told them to look at a specific marker. The instruction had two seconds of delay between one direction and another. The total duration was three minutes. We placed different markers – ‘Left’, ‘Right’, ‘Top’, ‘Bottom’, ‘Front’ in respective directions in the lab. We observed the participants' head movements and images rendered to ensure the participants were following the instructions. This simple task only required participants to look at the markers as directed. We wanted all the participants to observe the lab in a controlled fashion to make it consistent for all participants. The participants were standing straight on the balance board and not allowed to move their bodies, except for their heads, while exploring the VE. We collected real-time balance data from BTrackS Balance Plate.  Fig. 2 shows the real environment and corresponding VE. We replicated the task presented in \cite{ferdous2018investigating} to develop the standing visual exploration task. We chose to perform this motor task in a laboratory VE because we wanted to compare their balance in the VE lab to their balance in the real lab.

\paragraph{Standing Reach and Grasp Task:} Participants reached for and grasped real objects within their reach. We placed four objects (cube - 5.08 cm width) on the marked places on the table. The distance between every two objects was 24 cm. The balance board was placed on the ground in alignment with the middle of the table. The distance from the table to the balance board was 12 cm. Participants were barefoot and positioned each foot on the marked places on the balance board. Participants rested their non-dominant hands on the upper thigh and used the dominant hand to reach and grasp the objects. We allowed participants to reach the objects by leaning forward to a maximum comfortable distance without lifting their heels off the balance board and standing straight. We instructed the participants to grasp the four objects in random order, lift the objects to chest level, and put them back in the same place. We followed \cite{cordova2014older} to implement this task. Fig. 3 shows the workspace and a comparison between the real environment and  VE for this task. We chose this motor task, because reaching is a crucial part of everyday activities, has been used for balance measurement, and is common in VR \cite{bolton2019motor, huang2015effects, tan2012anticipatory}.

\subsubsection{Baseline Measurements without VR} The participants stood on a BTrackS Balance Plate while they were supported by a harness to protect them from falling. Then we measured their balance for the standing visual exploration task and standing reach and grasp task with three trials each. Each trial was three minutes long.

\subsubsection{VR Tasks} These are all replications of the above baseline tasks, except they were performed in VR, and thus there were subtle differences. Participants used the HTC Vive HMD to observe the VE for both virtual tasks. We repeated the following tasks for four different auditory conditions and a no audio VR condition with three trials for each session while the auditory conditions and tasks were counterbalanced.

\paragraph{Standing Visual Exploration Task in VR:} Participants followed the same recorded instructions as the one used for real-world balance measurement to explore the VE. We placed the virtual markers in the same location and the same size as the baseline measurement. The measurements were the same as in the baseline standing visual exploration task.

\paragraph{Standing Reach and Grasp Task in VR:}  Participants reached for and grasped virtual objects within their reach with their dominant hand using the controllers. When the participants touched the virtual objects with the controller, the object's color changed to red. Then the participants pulled the controller trigger to grasp the objects, brought them to chest level, and released the trigger when they returned the object to the same place. The virtual environment and measurements were the same as the baseline task.

\subsection{Post-Session Questionnaires}
Finally, the participants filled out an SSQ and demographic questionnaire at the end of the study. 

The whole study took around two hours for the participants to complete. Each participant received a payment of 30 US dollars per hour and money for parking fees at the end of the study.

\begin{figure}{}
\centering
\includegraphics[width=8.25cm,height=4cm]{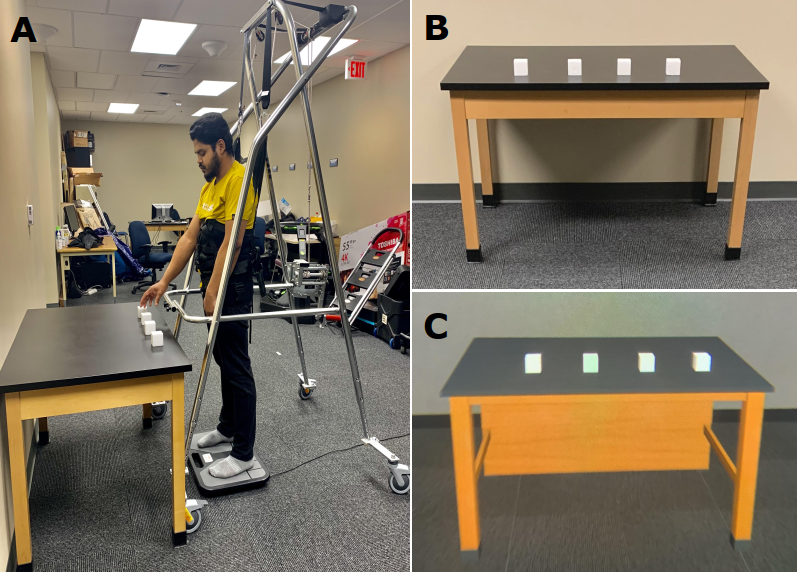}
\caption{Standing reach and grasp task: (A) Workspace (B) Real environment (C) Virtual environment}
\end{figure}


\section{Metrics}
\subsection{Center of Pressure (CoP) Velocity}
Center of Pressure (CoP) \cite{salavati2009test} velocity is the primary metric of balance in our study. We chose to use CoP velocity because it is widely used as a valid measurement to assess balance \cite{li2016reliability}. We calculated CoP from the four pressure sensors of the BTrackS Balance Plate, applying the following formula by Young et al. \cite{young2011assessing}:

\begin{equation}
CoP (X,Y) = \frac{\sum_{i=1}^{4} Weight_{i} * (x_{i},y_{i})} {\sum_{i=1}^{4} Weight_{i}} 
\end{equation}

Where $(x_{i},y_{i})$ = coordinates of the pressure sensor \textit{i}, $Weight_{i}$ = weight or pressure data on the $i$th sensor, and $CoP(X,Y)$ = coordinates of the $CoP$.

Then, we computed the CoP path for all samples using the following formula.

\begin{equation}
CoP\;Path = \sum_{i=1}^{n-1}\sqrt{(CoP_{i+1}X - CoP_{i}X)^2 + (CoP_{i+1}Y - CoP_{i}Y)^2}
\end{equation}

Here, $CoP_{i}X$ = $X$ coordinate of CoP at $i$th second, and $CoP_{i}y$ = $Y$ coordinate of CoP at $i$th second.

Finally, we calculated CoP velocity by dividing the CoP path for all samples by total data recording time for all samples (T).

\begin{equation}
    CoP\;Velocity = \frac{CoP\;Path}{T}
\end{equation}

\subsection{Activities-specific Balance Confidence (ABC) Scale}
ABC is a 16 item questionnaire where each item asks about participant's confidence in doing a specific daily life activity \cite{powell1995activities}. ABC score is calculated by the sum of the percentages from each question (1-16) for a maximum total of 1600. The sum is then divided by 16 to give the ABC\%.

\subsection{Simulator Sickness Questionnaire}
Simulator Sickness Questionnaire (SSQ) is a 16 item questionnaire where each item asks about participant's physiological discomfort \cite{kennedy1993simulator}. This measure is needed to identify participants who are prone to severe cybersickness and investigate the correlation with postural instability.

\section{Statistical Analysis}
 We used the Shapiro-Wilk test for checking data normality. We found the data was normally distributed for participants with and without BI for both tasks; $p$ = .321, $w$ = 0.89. Then to find any significant difference in CoP velocities, we performed a 2$\times$6 mixed-model ANOVA where we had two between-subject factors (participants with BI and participants without BI) and six within-subject factors (six study conditions: baseline, spatial, static, rhythmic, CoP, and no audio). When there was a significant difference, we conducted post-hoc two-tailed t-tests for within and between-group comparisons. For analyzing cybersickness, we also used two-tailed t-tests between pre-session SSQ score and post-session SSQ score for both groups of participants separately. We also performed two-tailed t-tests between the ABC score of both groups of participants to evaluate the difference in physical ability. We applied Bonferroni correction for all tests that involved multiple comparisons.

\section{Results}
We compared CoP velocities between study conditions and obtained the following results.
\subsection{Within Group Comparisons on CoP Velocity}
After performing ANOVA tests, we found a significant difference for participants with BI, \textit{F}(5,120) = 21.4, \textit{p} $<$ .001 for standing visual exploration task and \textit{F}(5,120) = 46.07, \textit{p} $<$ .001 for standing reach and grasp task. We also obtained a significant difference for participants without BI, \textit{F}(5,120) = 21.64, \textit{p} $<$ .001 for standing visual exploration task and \textit{F}(5,120) = 39.13, \textit{p} $<$ .001 for standing reach and grasp task. Then we performed the following pair-wise comparisons using two-tailed t-tests for both groups separately to find differences between particular study conditions.

\subsubsection{Baseline vs. No Audio}
\paragraph{Standing Visual Exploration Task:}
Experiment results did not show any significant difference between no audio (Mean, \textit{M} = 4.67, Standard Deviation, \textit{SD} = 1.63) and baseline (\textit{M}= 4.38, \textit{SD} = 1.57) condition; \textit{t}(20) = 1.72, \textit{p} = .101, \textit{r} = 0.88 for participants with BI. 
Similarly, we did not observe a significant difference between no audio (\textit{M} = 4.08, \textit{SD} = 1.33) and baseline (\textit{M}= 3.89, \textit{SD} = 1.49) condition; \textit{t}(20) = 1.11, \textit{p} = .279, \textit{r} = 0.86 for participants without BI.

\paragraph{Standing Reach and Grasp Task:}
We did not obtain any significant difference between no audio (\textit{M} = 6.44, \textit{SD} = 1.48) and baseline (\textit{M}= 6.21, \textit{SD} = 1.14) condition; \textit{t}(20) = 1.14, \textit{p} = .267, \textit{r} = 0.80 for participants with BI.
We also did not find a significant difference between no audio (\textit{M} = 5.63, \textit{SD} = 1.18) and baseline (\textit{M}= 5.27, \textit{SD} = 0.97) condition; \textit{t}(20) = 2.43, \textit{p} = .024, \textit{r} = 0.82 for participants without BI.
Although we expected a significant increase of CoP velocity in no audio in VR than the baseline condition, that did not happen in this case. 


\begin{figure}[ht!]
    \centering
  \includegraphics[width=0.5\textwidth]{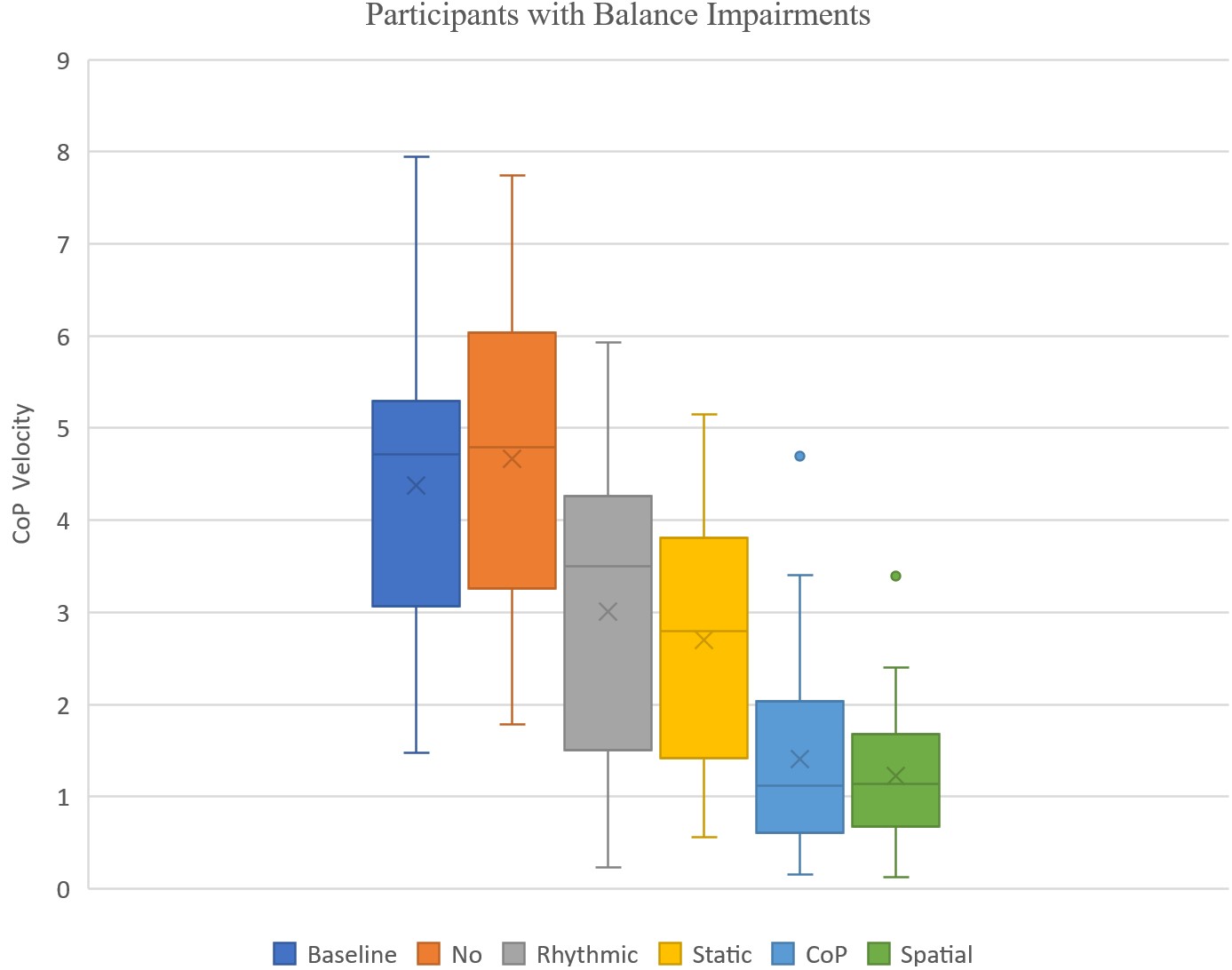}
   \includegraphics[width=0.5\textwidth]{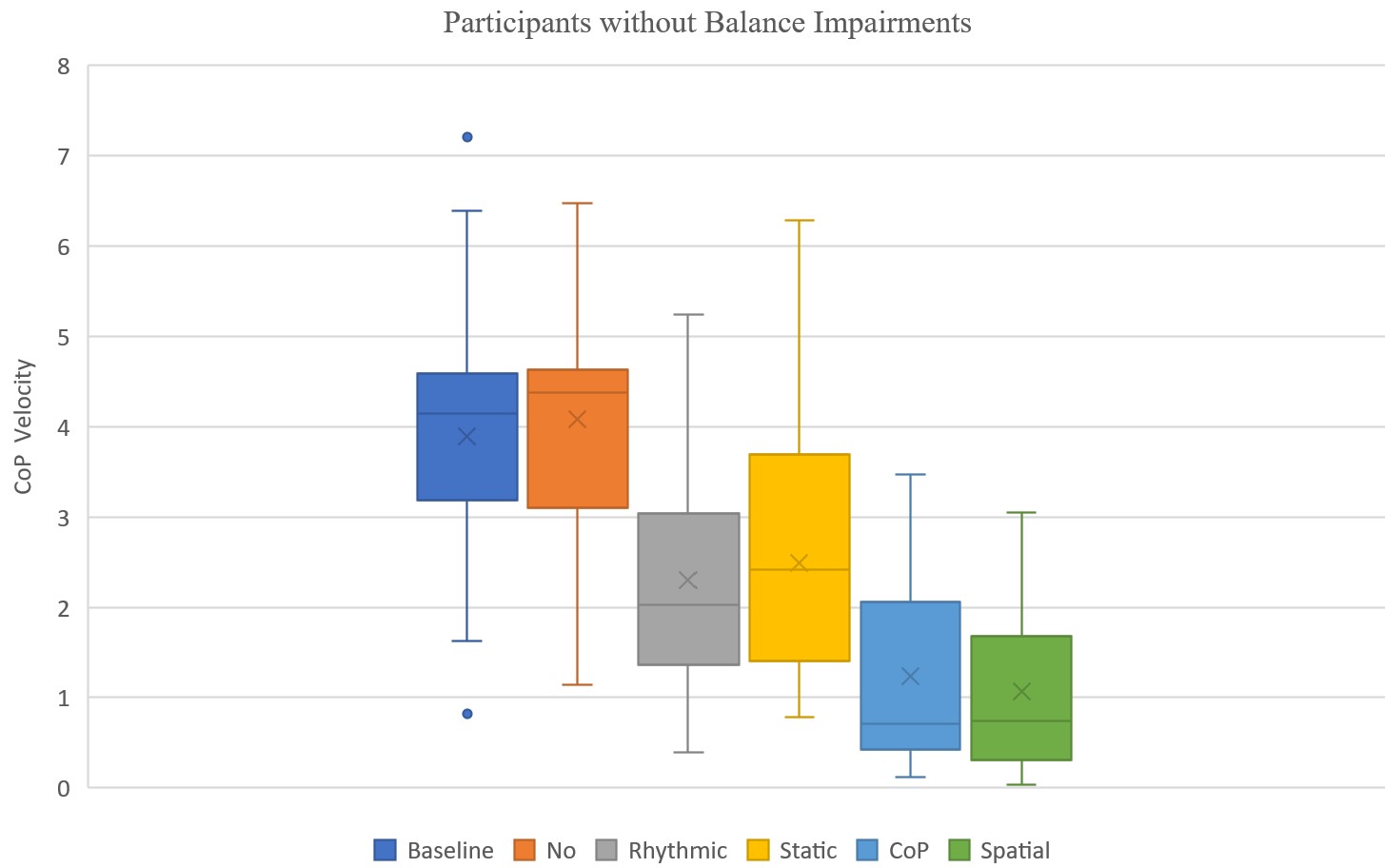}
  \caption{CoP velocity comparison between study conditions for standing visual exploration task.}
\end{figure}

\subsubsection{ No Audio vs. Spatial Audio}
\paragraph{Standing Visual Exploration Task:}
For participants with BI, CoP velocity was substantially lower in the spatial condition (\textit{M} = 1.22, \textit{SD} = 0.84 ) than no audio condition; \textit{t}(20) = 12.17, \textit{p} $<$ .001, \textit{r} = 0.61.
For participants without BI, we also observed in spatial condition (\textit{M} = 1.07, \textit{SD} = 0.95) CoP velocity was significantly lower than no audio condition; \textit{t}(20) = 10.02, \textit{p} $<$ .001, \textit{r} = 0.30. 

\paragraph{Standing Reach and Grasp Task:}
For participants with BI, the obtained CoP velocity was significantly less in spatial (\textit{M} = 2.07, \textit{SD} = 0.92 ) than no audio condition; \textit{t}(20) = 12.72, \textit{p} $<$ .001, \textit{r} = 0.22.
For participants without BI, we also found CoP velocity was significantly less in spatial (\textit{M} = 1.94, \textit{SD} = 0.88) than no audio condition; \textit{t}(20) = 11.71, \textit{p} $<$ .001, \textit{r} = 0.04.
Thus, spatial audio performed better than no audio in VR for both standing visual exploration and standing reach and grasp tasks.

\subsubsection{ No Audio vs. CoP Audio}
\paragraph{Standing Visual Exploration Task:}
CoP condition (\textit{M} = 1.41, \textit{SD} = 1.15) improved balance (i.e., decreased CoP velocity) significantly than no audio condition; \textit{t}(20) = 10.85, \textit{p} $<$ .001, \textit{r} = 0.56 for participants with BI. 
We also observed CoP velocity was significantly lower in CoP audio (\textit{M} = 1.23, \textit{SD} = 0.97 ) than no audio condition; \textit{t}(20) = 10.38, \textit{p} $<$ .001, \textit{r} = 0.44 for participants without BI.

\paragraph{Standing Reach and Grasp Task:}
The obtained CoP velocity was considerably less in CoP (\textit{M} = 2.27, \textit{SD} = 1.16) than no audio condition; \textit{t}(20) = 9.58, \textit{p} $<$ .001, \textit{r} = -0.12 for participants with BI.
We also found CoP velocity was significantly less in CoP audio (\textit{M} = 2.15, \textit{SD} = 0.93 ) than no audio condition; \textit{t}(20) = 10.42, \textit{p} $<$ .001, \textit{r} = -0.04 for participants without BI.
Therefore, the CoP audio condition outperformed no audio in VR condition for both tasks.

\subsubsection{ No Audio vs. Static Audio}
\paragraph{Standing Visual Exploration Task:}
For participants with BI, experimental results disclosed that CoP velocity was significantly less in static (\textit{M} = 2.70, \textit{SD} = 1.40 ) than no audio condition; \textit{t}(20) = 8.19, \textit{p} $<$ .001, \textit{r} = 0.75. 
For participants without BI, we also observed CoP velocity was significantly less in static audio (\textit{M} = 2.49, \textit{SD} = 1.36 ) than no audio condition; \textit{t}(20) = 7.30, \textit{p} $<$ .001, \textit{r} = 0.73.

\paragraph{Standing Reach and Grasp Task:}
The obtained CoP velocity was significantly less in static (\textit{M} = 4.24, \textit{SD} = 1.22) than no audio condition; \textit{t}(20) = 6.36, \textit{p} $<$ .001, \textit{r} = 0.33 for participants with BI.
We also discovered that CoP velocity was substantially lower in static audio (\textit{M} = 3.46, \textit{SD} = 1.29) than no audio condition; \textit{t}(20) = 6.52, \textit{p} $<$ .001, \textit{r} = 0.24 for participants without BI.
Hence, static rest frame audio also performed better than the no audio in VR condition for both tasks.

\subsubsection{ No Audio vs. Rhythmic Audio}
\paragraph{Standing Visual Exploration Task:}
Experiment results revealed that CoP velocity was significantly less in rhythmic (\textit{M} = 3.01, \textit{SD} = 1.77) than no audio condition for participants with BI; \textit{t}(20) = 10.77, \textit{p} $<$ .001, \textit{r} = 0.92. 
We also noticed that CoP velocity was significantly decreased in rhythmic audio (\textit{M} = 2.30, \textit{SD} = 1.33) than no audio condition for participants without BI; \textit{t}(20) = 6.99, \textit{p} $<$ .001, \textit{r} = 0.62.

\paragraph{Standing Reach and Grasp Task:}
The obtained CoP velocity was significantly reduced in rhythmic (\textit{M} = 4.10, \textit{SD} = 1.51) than no audio condition for participants with BI; \textit{t}(20) = 6.80, \textit{p} $<$ .001, \textit{r} = 0.45.
We also found CoP velocity was significantly diminished in rhythmic audio (\textit{M} = 3.33, \textit{SD} = 1.41 ) than no audio condition for participants without BI; \textit{t}(20) = 6.27, \textit{p} $<$ .001, \textit{r} = 0.17. 
Thus, it can be claimed that rhythmic audio outperformed no audio in VR condition.

\subsubsection{ Rhythmic Audio vs. Spatial Audio}
\paragraph{Standing Visual Exploration Task:}
We found that CoP velocity significantly decreased in spatial  than rhythmic audio condition for both participants with BI (\textit{t}(20) = 5.38, \textit{p} $<$ .001, \textit{r} = 0.52) and for participants without BI (\textit{t}(20) = 4.09, \textit{p} $<$ .001, \textit{r} = 0.31). 

\paragraph{Standing Reach and Grasp Task:}
From mixed ANOVA and post-hoc two-tailed t-test we noticed that CoP velocity was significantly less in spatial than rhythmic audio condition for participants with BI (\textit{t}(20) = 6.41, \textit{p} $<$ .001, \textit{r} = 0.37) and for participants without BI (\textit{t}(20) = 5.53, \textit{p} $<$ .001, \textit{r} = 0.58).
These results indicated that spatial audio could be preferred over rhythmic audio for balance improvement.

\begin{figure}[ht!]
    \centering
  \includegraphics[width=0.5\textwidth]{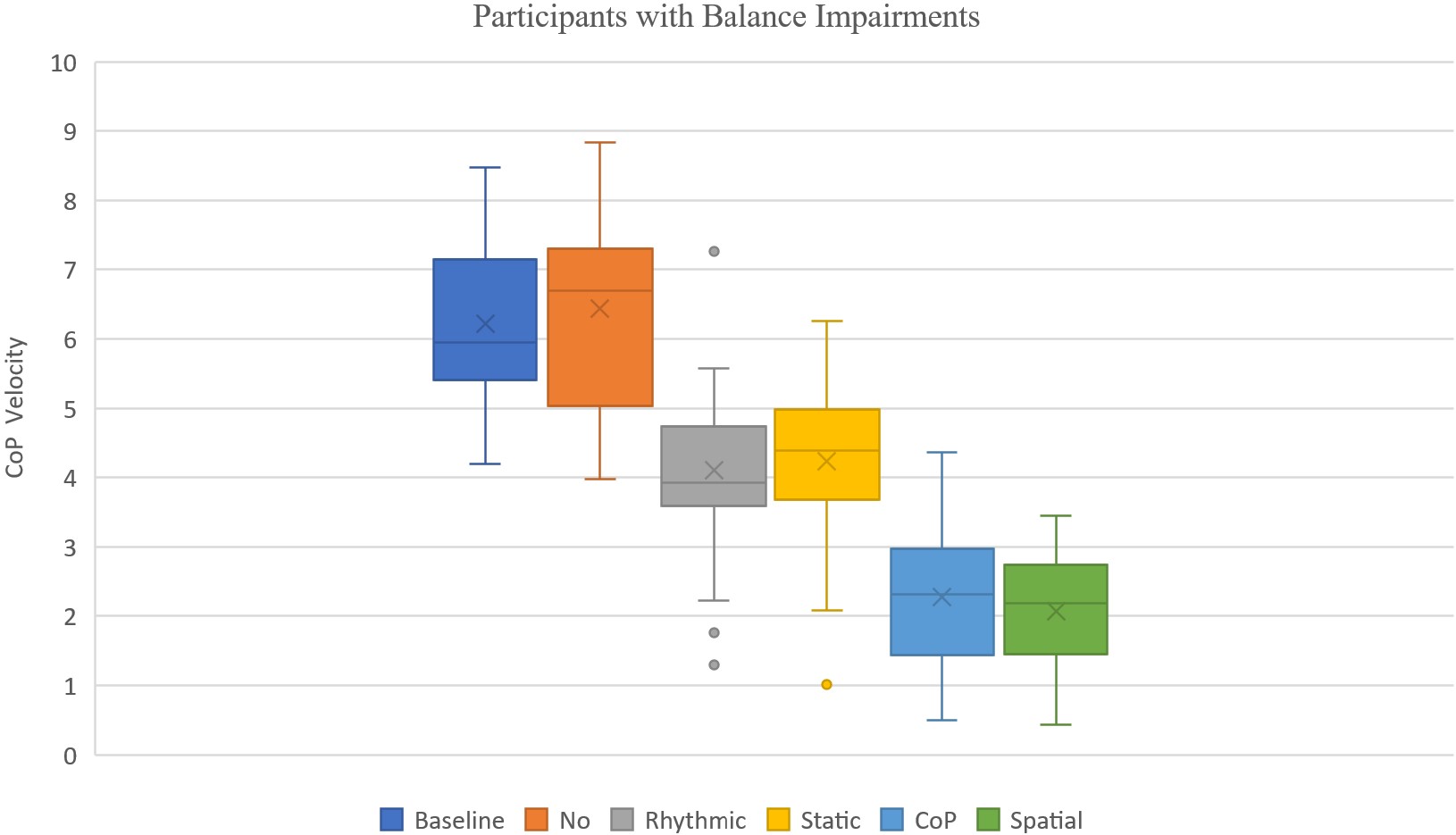}
  \includegraphics[width=0.5\textwidth]{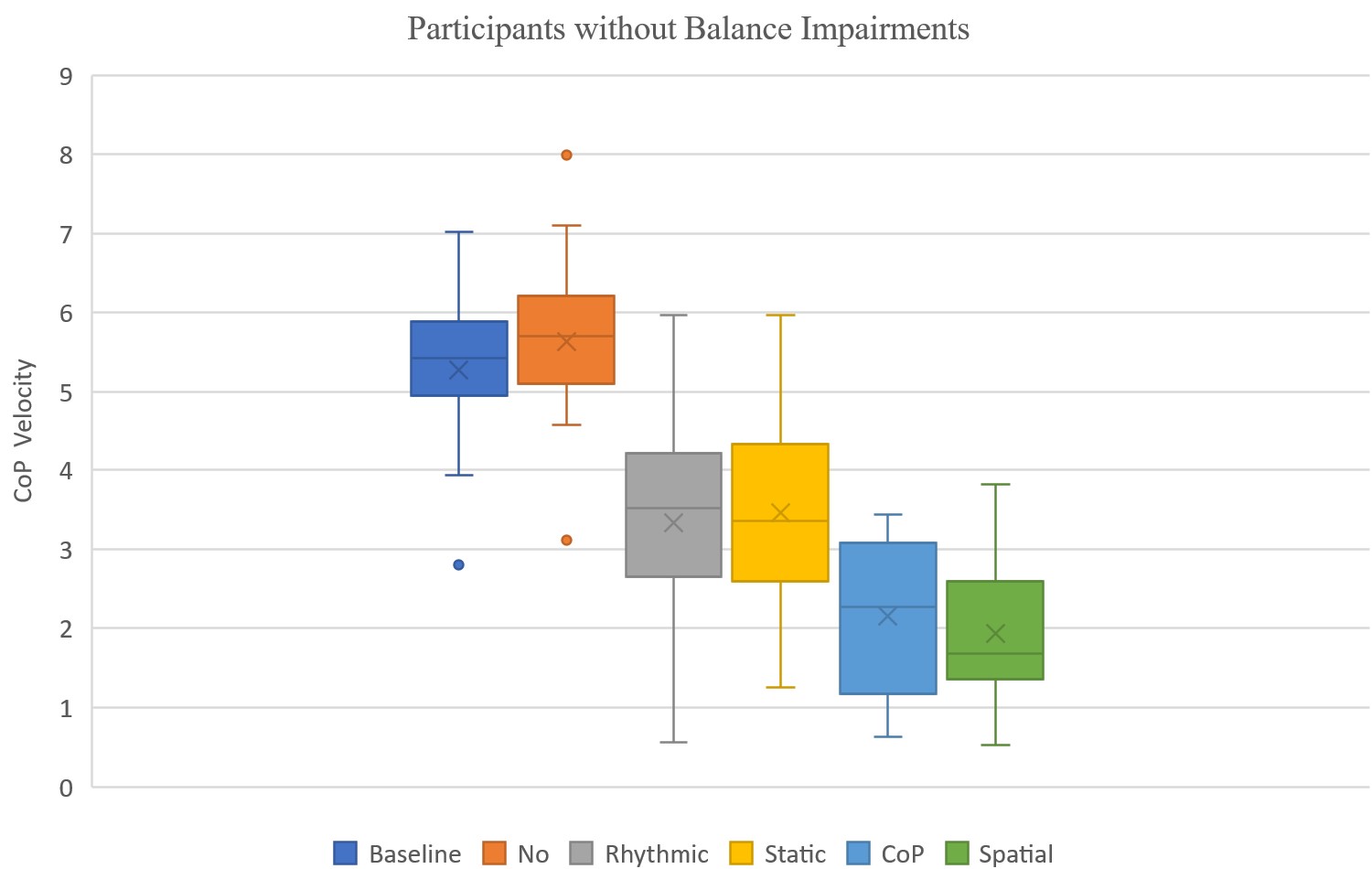}
  \caption{CoP velocity comparison between study conditions for standing reach and grasp task.}
\end{figure}

\subsubsection{ Rhythmic Audio vs. CoP Audio}
\paragraph{Standing Visual Exploration Task:}
Experiment results substantiated that CoP velocity was significantly less in CoP audio than rhythmic audio condition for participants with BI (\textit{t}(20) = 4.68, \textit{p} $<$ .001, \textit{r} = 0.49) and for participants without BI (\textit{t}(20) = 5.12, \textit{p} $<$ .001, \textit{r} = 0.70).

\paragraph{Standing Reach and Grasp Task:}
The obtained CoP velocity was significantly lower in CoP than rhythmic audio condition for participants with BI (\textit{t}(20) = 4.73, \textit{p} $<$ .001, \textit{r} = 0.14) and for participants without BI (\textit{t}(20) = 5.15, \textit{p} $<$ .001, \textit{r} = 0.67). The outcomes supported that CoP audio might be better than rhythmic audio for improving balance in VR.

\subsubsection{ Rhythmic Audio vs. Static Audio}
\paragraph{Standing Visual Exploration Task:}
Experiment results revealed that there is no significant difference between rhythmic audio and static audio condition for participants with BI (\textit{t}(20) = 1.24, \textit{p} = .229, \textit{r} = 0.77) and for participants without BI (\textit{t}(20) = 0.99, \textit{p} = .333, \textit{r} = 0.78).

\paragraph{Standing Reach and Grasp Task:}
We also did not obtain a significant difference between rhythmic audio and static audio condition for participants with BI (\textit{t}(20) = 0.64, \textit{p} = .532, \textit{r} = 0.77) and for participants without BI (\textit{t}(20) = 0.97, \textit{p} = .344, \textit{r} = 0.90).
Thus, the results did not clearly indicate which audio could be chosen between rhythmic and static rest frame for balance improvement.

\subsubsection{ Static Audio vs. Spatial Audio}
\paragraph{Standing Visual Exploration Task:}
Experiment results substantiated that CoP velocity was significantly lower in spatial audio than static audio condition for participants with BI (\textit{t}(20) = 6.04, \textit{p} $<$ .001, \textit{r} = 0.60) and for participants without BI (\textit{t}(20) = 5.69, \textit{p} $<$ .001, \textit{r} = 0.56).

\paragraph{Standing Reach and Grasp Task:}
The obtained CoP velocity was significantly less in spatial than static audio condition for participants with BI (\textit{t}(20) = 8.31, \textit{p} $<$ .001, \textit{r} = 0.41). The same result was found for participants without BI (\textit{t}(20) = 6.48, \textit{p} $<$ .001, \textit{r} = 0.56).
As a result, spatial audio could be favored over static rest frame for balance improvement in VR.

\subsubsection{ Static Audio vs. CoP Audio}
\paragraph{Standing Visual Exploration Task:}
We discovered that CoP velocity was significantly less in CoP audio than static audio condition; \textit{t}(20) = 5.38, \textit{p} $<$ .001, \textit{r} = 0.64 for participants with BI and for participants without BI (\textit{t}(20) = 6.87, \textit{p} $<$ .001, \textit{r} = 0.79).

\paragraph{Standing Reach and Grasp Task:}
CoP velocity was significantly less in CoP than static audio condition; \textit{t}(20) = 6.04, \textit{p} $<$ .001, \textit{r} = 0.22 for participants with BI and for participants without BI (\textit{t}(20) = 5.49, \textit{p} $<$ .001, \textit{r} = 0.56).
Results displayed that CoP audio performed significantly better than static rest frame audio for both tasks.

\subsubsection{ CoP Audio vs. Spatial Audio}
\paragraph{Standing Visual Exploration Task:}
Experiment results did not show a significant difference between spatial audio and CoP audio condition for both participants with BI (\textit{t}(20) = 1.16, \textit{p} = .13, \textit{r} = 0.77) and participants without BI (\textit{t}(20) = 0.68, \textit{p} = .253, \textit{r} = 0.30).

\paragraph{Standing Reach and Grasp Task:}
We discovered similar results in this case as the standing visual exploration. We did not obtain a significant difference between spatial and CoP audio condition for participants with BI (\textit{t}(20) = 0.72, \textit{p} = .238, \textit{r} = 0.31) and for participants without BI (\textit{t}(20) = 0.97, \textit{p} = .172, \textit{r} = 0.39).
Therefore, the results did not indicate which audio can be preferred between spatial and CoP for improving balance in VR environments.

\begin{table}[t]
\caption{Summarized results for pairwise comparisons}
    \label{tab:my_label}
    \setlength{\tabcolsep}{1.5pt}
\begin{tabular}{|@{}l@{}|@{}cc@{}|@{}cc|}
\hline
\multicolumn{1}{|c|}{\multirow{0}{*}{}{\textbf{Comparisons }}} & \multicolumn{2}{c|}{\textbf{\begin{tabular}[c]{@{}c@{}}Standing Visual \\ Exploration\end{tabular}}} & \multicolumn{2}{c|}{\textbf{\begin{tabular}[c]{@{}c@{}}Standing Reach\\  and Grasp\end{tabular}}} \\ \cline{2-5} 
                                      & \multicolumn{1}{c|}{BI}                    & \begin{tabular}[c]{@{}c@{}}Without BI\end{tabular}   & \multicolumn{1}{c|}{BI}                  & \begin{tabular}[c]{@{}c@{}}Without BI\end{tabular}  \\ \hline
\textbf{Spatial vs. Static}           & \multicolumn{1}{c|}{p \textless .001}      & p \textless .001                                        & \multicolumn{1}{c|}{p \textless .001}    & p \textless .001                                       \\ \hline
\textbf{Spatial vs. Rhythmic}         & \multicolumn{1}{c|}{p \textless .001}      & p \textless .001                                        & \multicolumn{1}{c|}{p \textless .001}    & p \textless .001                                       \\ \hline
\textbf{Spatial vs. CoP}              & \multicolumn{1}{c|}{p \textless .001}      & p \textless .001                                        & \multicolumn{1}{c|}{p \textless .001}    & p \textless .001                                       \\ \hline
\textbf{Spatial vs. No audio}               & \multicolumn{1}{c|}{p \textless .001}      & p \textless .001                                        & \multicolumn{1}{c|}{p \textless .001}    & p \textless .001                                       \\ \hline
\textbf{Static vs. Rhythmic}          & \multicolumn{1}{c|}{p \textgreater .05}    & p \textgreater .05                                      & \multicolumn{1}{c|}{p \textgreater .05}  & p \textgreater .05                                     \\ \hline
\textbf{Static vs. CoP}               & \multicolumn{1}{c|}{p \textless .001}      & p \textless .001                                        & \multicolumn{1}{c|}{p \textless .001}    & p \textless .001                                       \\ \hline
\textbf{Static vs. No audio}                & \multicolumn{1}{c|}{p \textless .001}      & p \textless .001                                        & \multicolumn{1}{c|}{p \textless .001}    & p \textless .001                                       \\ \hline
\textbf{Rhythmic vs. CoP}             & \multicolumn{1}{c|}{p \textless .001}      & p \textless .001                                        & \multicolumn{1}{c|}{p \textless .001}    & p \textless .001                                       \\ \hline
\textbf{Rhythmic vs. No audio}              & \multicolumn{1}{c|}{p \textless .001}      & p \textless .001                                        & \multicolumn{1}{c|}{p \textless .001}    & p \textless .001                                       \\ \hline
\textbf{CoP vs. No audio}                   & \multicolumn{1}{c|}{p \textless .001}      & p \textless .001                                        & \multicolumn{1}{c|}{p \textless .001}    & p \textless .001                                       \\ \hline
\textbf{Baseline vs. No audio}              & \multicolumn{1}{c|}{p \textgreater .05}    & p \textgreater .05                                      & \multicolumn{1}{c|}{p \textgreater .05}  & p \textgreater .05                                     \\ \hline
\end{tabular}
\end{table}
Fig. 4 and Fig. 5 represents the experimental results for standing visual exploration task and standing reach and grasp task, respectively. Table 2 represents the summarized results.

\subsection{Between Group Comparisons}
Results from mixed-model ANOVA and post-hoc two-tailed t-tests indicated that there was a significant difference in CoP velocities for baseline conditions between the two groups (participants with and without BI); \textit{t}(20) = 8.31, \textit{p} $<$ .001, \textit{r} = 0.41.
However, we did not observe any significant difference between other study conditions.

\subsection{Activities-specific Balance Confidence (ABC) Scale}
We administered the Activities Specific Balance Scale (ABC-16) for both participants with and without BI, which can be interpreted as follows: 80\% = high level of physical functioning; 50-80\% = moderate level of physical functioning; $<$ 50\% = low level of physical functioning. We performed a two-tailed t-test between the ABC score of the participants with BI ($M$ = 70.83, $SD$ = 24.83) and those without BI ($M$ = 91.76, $SD$ = 13.71), \textit{t}(20) = 3.38, \textit{p} $<$ .001. The mean ABC score of the participants with BI was 70.83\%, indicating the participants with BI had a moderate level of physical functioning. On the other hand, the mean ABC score of the participants without BI was 91.76\%, confirming their high level of physical functioning. 

\subsection{Simulator Sickness Questionnaire}
We performed a two-tailed t-test between pre-session SSQ score and post-session SSQ score for both groups of participants with BI and those without BI. We did not find any significant difference between the pre-session SSQ score and post-session SSQ score for both groups of participants. We obtained \textit{t}(20) = 1.72, \textit{p} = .08, \textit{r} = 0.6 for participants with BI and \textit{t}(20) = 1.72, \textit{p} = .06, \textit{r} = 0.77 for participants without BI.

\section{Discussion}
\subsection{Effect of Auditory Conditions on Balance in VR}
Balance improves with the decrease of CoP velocity \cite{thompson2017balance,ruhe2011center}. We observed the following differences in balance from our experimental results.

\subsubsection{No Audio in VR vs. All Auditory Conditions in VR}
Experimental results showed that the CoP velocity was significantly lower in all audio conditions compared to the no audio in VR condition for both participants with and without BI for both tasks (standing visual exploration and standing reach and grasp). Thus, we can say that spatial, CoP, rhythmic, and static rest frame audio improved balance significantly for both the participants with and without BI. As a result, hypothesis 1 cannot be rejected. These results supported prior studies where they reported auditory white noise  \cite{sacco2018effects,zhou2021effects,ross2015auditory,harry2005balancing}, spatial \cite{stevens2016auditory,gandemer2017spatial}, CoP \cite{hasegawa2017learning}, static rest frame \cite{ross2016auditory}, and rhythmic audio \cite{ghai2018effect} improved balance in the real-world environments. 

However, the results indicated a difference compared to prior work \cite{ferdous2018investigating} where they investigated the effect of visual feedback on balance using the standing visual exploration task for both participants with and without BI in VR. They reported that the visual feedback improved balance for the people with BI, but they did not find any significant effects of any visual conditions on balance for the people without BI. However, our experimental results suggested that the auditory feedback improved balance for both participants with and without BI. We hypothesized from the results that the specific auditory feedback approaches evaluated might perform better than the visual feedback. Similar results were found in prior studies \cite{baram2012gait} where auditory feedback was more effective than visual feedback during walking in VR.



\subsubsection{ Comparison Between All Auditory Conditions in VR}
The experimental results demonstrated differences between the audio conditions. Results suggested that spatial and CoP audio conditions perform significantly better than the rhythmic and static rest frame audio conditions in both tasks of standing visual exploration and standing reach and grasp for both the participants with and without BI. However, there was no significant difference statistically between groups in the spatial and CoP audio conditions. The reason behind this could be that both the tasks analyzed were stationary tasks using a simple VE. Otherwise, spatial audio would have a greater impact on an individual in motion as spatial audio provides greater immersion and stabilization in an immersive VR environment \cite{wenzel2017perception,naef2002spatialized}. Thus, hypothesis 2 was only partially supported.

Furthermore, static rest frame audio performed somewhat better than rhythmic audio except in standing visual exploration tasks for participants without BI, where rhythmic audio outperformed static rest frame audio slightly. We hypothesized that this happened because the participants with BI were more focused on keeping their balance during the visual exploration task than maintaining synchronicity with the rhythmic audio beat. From the post-session comments of the participants, we observed that participants without BI were often able to effectively follow the command pattern of pre-recorded instruction (e.g., left, right, top, bottom, front) for standing visual exploration task which might have given them an advantage in contrast with balance-impaired individuals who might have had a little more fatigue as we noticed that they rested longer than the participants without BI in-between conditions. 
However, we did not obtain a significant difference between groups for rhythmic and static rest frame audio conditions. 

In our study, we discovered spatial audio outperformed all other auditory conditions. Prior work also found that spatial auditory feedback was the most preferred over other types as it provided greater fidelity \cite{chong2020audio, pinkl2020spatialized} and immersion \cite{wenzel2017perception,naef2002spatialized}. However, minimal prior work quantitatively compares spatial audio to other audio techniques with respect to the effects on balance in VR.

\subsection{Significant Effect of Auditory Changes with Position}
The auditory conditions improved balance in VR but to a differing amounts. The results demonstrated that spatial and CoP audio improved balance significantly more than other auditory conditions. Spatial and CoP audio were related to participant's position, i.e., when the participants leaned slightly in any direction while standing on the balance board or moved their heads, there was a change in audio volumes. As these two auditory conditions performed significantly better than other auditory conditions, we hypothesized that the audio volume that changes with participant's positioning provides significantly better feedback to the participants to correct their posture. However, spatial audio may be subjectively preferred over CoP audio as spatial audio provides greater immersion in VR \cite{wenzel2017perception,naef2002spatialized}.

\subsection{Between Group Comparisons}
The ABC scores indicated that participants with BI had potentially lower physical functionality than participants without BI. Thus, it was reasonable to have the difference in CoP velocities between baseline conditions for the two groups. However, there was no significant difference between other conditions in VR. For further investigation, we subtracted baseline data from all situations to see which group improved balance the most. Then, we performed a mixed ANOVA and post-hoc two-tailed t-tests across two groups which demonstrated that the participants with BI improved balance significantly more than those without BI. We anticipated that because the participants with BI had reduced balance functioning, they might have a better chance of improving balance in VR than those without BI. This result was also substantiated by the prior work where they found that participants with BI improved balance and gait significantly more than the participants without BI \cite{guo2015mobility}. We hypothesized that as the participants with BI improved balance significantly more than the participants without BI in VR, we did not get any significant difference between the two groups in VR even though there was a significant difference between baseline conditions.

\subsection{Effect of Virtual Environment}
We evaluated no audio condition in VR and the baseline condition without VR. We found the CoP velocity was slightly higher, albeit not significantly, in the no audio in VR condition compared to the baseline condition for both standing visual exploration and standing reach and grasp tasks for both groups of participants. Although significant differences were not found, prior research suggested that postural instability generally increases in VR \cite{epure2014effect,soffel2016postural}, leading to increased CoP velocity in VR than in a baseline condition. We suspected that if we recruited more participants, we would see significant differences in the no audio in VR and baseline conditions.

\subsection{Cybersickness}
We did not find a significant difference between the pre-session SSQ score and the post-session SSQ score for both participants with and without BI, which indicated that participants were not affected by cybersickness. Participants might have developed mild cybersickness as our study had two different tasks, each task had six different conditions, and each condition had three trials which made the study take almost two hours to complete. Cybersickness is common when engaged in VR activities for over 10 minutes \cite{chang2020virtual,kim2021clinical}. However, our environment was simple and designed to minimize cybersickness as there was no illusory self-motion \cite{mccauley1992cybersickness}. Therefore, we assumed that cybersickness did not affect CoP velocity results.

\subsection{Limitations}

Participants were informed about the whole study procedure at the beginning of the study. Then we had a few trials with them until they were comfortable with the experimental procedure. We also had three baseline trials for each of the two tasks before starting the auditory conditions in VR. Each baseline trial duration was three minutes. However, the baseline could be extended for more trials. We did not do that as the study took almost two hours to complete and included participants with balance impairments who had potentially lower physical functionality. However, we counterbalanced both tasks and the auditory conditions in VR to mitigate the learning effects.

For designing our CoP audio condition, we streamed the CoPx and CoPy data from the balance board to Unity through sockets and mapped the CoPx to pitch and CoPy to stereo pan for providing participants CoP audio feedback according to their positioning on the balance board. However, it is unclear how our results for this condition would be affected at lower levels of latency. While we did not measure end-to-end latency, the additional network latency in this condition was negligible.

We did not adjust the table height based on the participant's height for the standing reach and grasp task, which could have affected results. However, there were no significant differences in the heights of the participants (Table 1) and thus, we expected it had a minimal effect.

Participants used harnesses for the whole study to protect themselves from falling, which might have improved their balance slightly. However, to make the study procedure consistent and safe, we required all participants to use the harnesses regardless of whether they had balance impairments or not. Thus, studies looking at balance not using a harness may have different results.

The study duration was quite long and required participants to hear the white noise standing in one place. This often caused fatigue, and individuals had to rest for a few minutes, taking the HMD off between different trials. This might have allowed participants to regain spatial awareness and regain balance which might have skewed data.

We calculated mean CoP velocity, which is a very popular metric for balance measurement \cite{li2016reliability}. However, we did not measure full-body movement. 

Due to COVID-19 and our targeted test population, primarily persons with BI due to MS, the recruitment process was difficult as many potential participants had compromised immune systems making them at high risk for COVID-19. Thus, they did not participate in the study. If the study had been done outside of COVID-19, we would have been able to recruit more participants. In that case, we might have found a significant difference between no audio in VR and the baseline for within group comparisons. We also might have found more significant differences among different study conditions in VR for between group comparisons. More research is needed to confirm this. 

\section{Design Implications}
Spatial and CoP auditory feedback performed significantly better than rhythmic and static rest frame audio conditions. Thus, spatial and CoP audio can be used in future VR environments to improve balance for VR users, especially for participants with balance impairments. However, spatial audio might be favored over CoP audio, as spatial audio renders greater immersion  \cite{wenzel2017perception,naef2002spatialized} and fidelity \cite{chong2020audio, pinkl2020spatialized} in VR. Hypothetically, audio techniques may interfere with presence less than visual techniques, for example,  because most VEs are predominantly visual experiences. In the future, we plan to investigate how different modalities of feedback, e.g., visual, auditory, tactile, affect balance in VR. Also, end-to-end latency might need to be considered, which we did not measure in our study.

\section{Conclusion}
In this paper, we evaluated the effect of different auditory feedback techniques on balance in VR. All auditory conditions (spatial, CoP, rhythmic, and static rest frame) improved balance significantly in our study. Spatial and CoP audio significantly outperformed rhythmic and static rest frame audio. However, there was no significant difference between spatial and CoP audio or between rhythmic and static rest frame audio. The results will help researchers better understand the different kinds of auditory feedback for maintaining balance in an HMD-based VE. Moreover, this research can help developers create VR experiences that are more usable and accessible to persons with and without balance impairments. In our future work, we will include locomotion tasks and investigate the effectiveness of auditory feedback for gait improvement.

\acknowledgments{
This work was funded through a grant from the National Science Foundation (IIS 2007041). We would also like to thank Dr. Sharif Mohammad Shahnewaz Ferdous for his directions towards developing the standing visual exploration scene.}

\bibliographystyle{abbrv-doi}

\bibliography{template}
\end{document}